\documentclass[prl,showpacs,superscriptaddress,preprintnumbers,amsmath]{revtex4}
\usepackage[dvips,final]{graphicx}
  \usepackage{amssymb}
   \usepackage{amsmath}
    \usepackage{epsfig}
     \usepackage{bm}
      \usepackage{pifont}
\begin{document}
\preprint{\hbox{\tt RUB-TPII-10/2010}}

\title{Taming singularities in transverse-momentum-dependent parton densities\footnote{Talk presented by the first author at the International Workshop
on Diffraction in High-Energy Physics, 10-15 Sept, 2010, Otranto (Lecce), Italy
}}


\author{I.~O.~Cherednikov}
\email{igor.cherednikov@jinr.ru}
\affiliation{Bogoliubov Laboratory of Theoretical Physics,
             Joint Institute for Nuclear Research,
             RU-141980 Dubna, Russia}
\affiliation{INFN Gruppo collegato di Cosenza,
             Dipartimento di Fisica, Universit$\grave{a}$
             della Calabria, I-87036 Arcavacata di Rende (CS), Italy}
\affiliation{ITPM,
             Moscow State University, RU-119899, Moscow, Russia}
\author{N.~G.~Stefanis}
\email{stefanis@tp2.ruhr-uni-bochum.de}
\affiliation{Institut f\"{u}r Theoretische Physik II, \\
             Ruhr-Universit\"{a}t Bochum,
             D-44780 Bochum, Germany}

\begin{abstract}
 We propose a consistent treatment of divergences emerging in the
 computation of transverse-momentum-dependent parton densities in
 leading $\alpha_s$-order of QCD perturbation theory.
\end{abstract}

\maketitle

Transverse-momentum-dependent (TMD) parton distribution functions (PDF)s
(abbreviated in what follows by the term ``TMD'')
accumulate information about the intrinsic 3-dimensional motion of
partons in a hadron \cite{S79}.
They depend, therefore, on the longitudinal ($x = k^+/p^+$), as well
as on the transverse $({\bm k}_\perp)$ momentum fractions of a
given parton.
Trying to work out a consistent operator definition of TMDs, one
encounters the puzzle of emergent divergences \cite{CS82, Col08}.
These, being hidden in the case of collinear PDFs, become visible in
the TMDs and jeopardize, in particular, their renormalizability
\cite{CS07, CS08, CS09, SC09, CKS10}.
In the present work, we explore the issue of extra rapidity divergences
in the TMDs in leading $\alpha_s$-order and describe a consistent
method to take care of them.

We start from the definition of a TMD (of a quark with flavor $i$ in
a hadron $h$) that respects gauge invariance and collinear factorization
on the tree-level \cite{CSS89, Col03, JY02, BJY03, BMP03, BR05},
but has no concern with any singularities --- as these arise only in the
one-loop corrections:
\begin{eqnarray}
  && {\cal F}_{i/h}^{\rm tree} \left(x, {\bm k}_\perp\right)
=
  \frac{1}{2}
  \int \frac{d\xi^- d^2{\bm \xi}_\perp}{2\pi (2\pi)^2}
  {\rm e}^{-ik^{+}\xi^{-} + i {\bm k}_\perp
\cdot {\bm \xi}_\perp}
  \left\langle
              h |\bar \psi_i (\xi^-, {\bm \xi}_\perp)
              [\xi^-, {\bm \xi}_\perp;
   \infty^-, {\bm \xi}_\perp]^\dagger
\right. \nonumber \\
&& \left. \times
   [\infty^-, {\bm \xi}_\perp;
   \infty^-, {\bm \infty}_\perp]^\dagger
   \gamma^+[\infty^-, {\bm \infty}_\perp;
   \infty^-, \mathbf{0}_\perp]
   [\infty^-, \mathbf{0}_\perp; 0^-,\mathbf{0}_\perp]
   \psi_i (0^-,\mathbf{0}_\perp) | h
   \right\rangle \ \, .
\label{eq:tmd_naive}
\end{eqnarray}
Tree-level gauge invariance is ensured by the inserted gauge links
(path-ordered Wilson-line operators) having the generic form
\begin{equation}
  { [y,x|\Gamma] }
=
  {\cal P} \exp
  \left[-ig\int_{x[\Gamma]}^{y}dz_{\mu} {\cal A}^\mu (z)
  \right] \ ,
\label{eq:link}
\end{equation}
where ${\cal A} \equiv t^a A^a$.
The transverse gauge links, extending to light-cone infinity
\cite{JY02, BJY03, BMP03}, are also included in (\ref{eq:tmd_naive}).
Beyond the tree level, the function (\ref{eq:tmd_naive}) will be shown
to be dependent on the renormalization scale $\mu$ and the rapidity
cutoff $\eta$.
We assume that any soft and collinear singularities can be properly
factorized out and be treated by means of the standard procedure
so that we don't have to consider them anymore.
Thus, we only concentrate on the unusual divergences which are
specific to the TMD case.


It was shown in \cite{CS08} that in the light-cone gauge, the
anomalous divergent term containing overlapping
(${\rm UV} \otimes {\rm rapidity}$ singularity) stems from the
virtual-gluon contribution
\begin{equation}
     \Sigma^{\rm LC}_{\rm virt.}
     =
     - \frac{\alpha_s}{\pi} C_{\rm F} \  \Gamma(\epsilon)\
  \left[ 4 \pi \frac{\mu^2}{-p^2} \right]^\epsilon\
  \delta (1-x) \delta^{(2)} (\bm k_\perp)\ \int_0^1\!
  dx \frac{(1-x)^{1-\epsilon}}{x^\epsilon [x]_\eta} \ ,
\label{eq:sigma-lc}
\end{equation}
where the UV divergence is treated within the dimensional-regularization
$\omega = 4 - 2\epsilon$ approach, while the rapidity divergence in the
gluon propagator in the light-cone gauge is regularized by the parameter
$\eta$, entailing the following regularization of the last integral in
Eq.\ (\ref{eq:sigma-lc}):
\begin{equation}
   \frac{1}{[x]_{\eta}^{\rm Ret. / Adv. / P.V.} }
   =  \left[ \frac{1}{x + i \eta}\ , \quad\quad
   \frac{1}{x - i \eta}\ , \quad\quad
   \frac{1}{2}\left(\frac{1}{x+i \eta}
                             +\frac{1}{x - i \eta}\right) \right]\ .
\end{equation}
Within this approach, one can extract from Eq.\ (\ref{eq:sigma-lc})
the UV-divergent part and obtain the overlapping singularity in the
logarithmic form
\begin{equation}
 {\Sigma}^{\rm LC}_{\rm virt.}
 =
   {
   - \frac{\alpha_s}{\pi}C_{\rm F} \   \frac{1}{\epsilon}
   \left[- \frac{3}{4} -  \ln \frac{\eta}{p^+} + \frac{i\pi}{2}
    \right] } + [{\rm UV \ finite\ part}]\ ,
\end{equation}
where the contribution of the transverse link is taken into account,
while the mirror diagram is omitted (see for technical details in
\cite{CS08}).
The exact form of the overlapping singularity drops us a hint at the
form of the additional soft factor which must be introduced into the
definition of TMD (\ref{eq:tmd_naive}), if one wants to extend it
beyond the tree level in order to render it renormalizable and free
of undesirable divergences --- at least at one-loop \cite{CS07, CS08}.
Hence, a generalized renormalization procedure has been formulated
\cite{KR87} in terms of a soft factor supplementing the tree-level
TMD, i.e.,
\begin{equation}
  {\cal F}^{\rm tree}(x, {\bm k}_\perp)
\to
  {\cal F} (x, {\bm k}_\perp; \mu, \eta, \epsilon) \times R^{-1}
(\mu, \eta, \epsilon) \ ,
\end{equation}
so that the above expression is free of overlapping divergences and can
be renormalized by means of the standard $R-$operation.
Within this framework, the introduction of the small parameter $\eta$
allows one to keep the overlapping singularities under control and
treat the extra term in the UV-divergent part via the cusp anomalous
dimension, which in turn determines the specific form of the gauge
contour in the soft factor $R$.

It is worth comparing the result obtained in the light-cone gauge with
the calculation in covariant gauges.
In Ref.\ \cite{CS82}, it was shown that the virtual-gluon exchange
between the quark line and the light-like gauge link
(this graph is obviously absent in the light-cone gauge) yields
(in the dimensional regularization)
\begin{equation}
 \Sigma_{\rm virt.}^{\rm cov.}
 =
 - \frac{\alpha_s}{\pi} C_{\rm F} \Gamma (\epsilon) \left[4\pi \frac{\mu^2}{-p^2} \right]^{\epsilon} \
 \delta (1-x) \delta^{(2)} (\bm k_\perp)\ \int_0^1\! dx \ \frac{x^{1-\epsilon}}{(1-x)^{1+\epsilon}} \ .
 \label{eq:sigma-cov}
\end{equation}
This expression contains the double pole $1/\epsilon^2$, which is not
compensated by the real counter part in the TMD case, while in collinear
PDFs, such a compensation does indeed take place.
Going back to our Eq.\ (\ref{eq:sigma-lc}), we observe that without
the $\eta$-regularization of the last integral and after a trivial change
of variables it is reduced to Eq.\ (\ref{eq:sigma-cov}) and reads
\begin{equation}
  \Sigma_{\rm virt.}^{\rm cov.} (\epsilon)
  =
  \Sigma_{\rm virt.}^{\rm LC} (\epsilon, \eta = 0) \ .
\end{equation}
The latter result allows us to conclude that the generalized
renormalization procedure, described above, is gauge invariant and
regularization-independent: in principle, one can use dimensional
regularization to take care of the overlapping singularities as well.
However, in the latter case the structure of extra divergences is much
less transparent and one is not able to conclude about the specific
form of the soft factor.
Let us note that the applicability of dimensional regularization to
for a consistent treatment of the divergences arising in the path-dependent
gauge invariant two-quark correlation function had been studied in
Ref.\ \cite{Ste83}.

Another obstacle still arises in the soft factor.
Evaluating in the light-cone gauge one-loop graphs, one finds the
expression
\begin{equation}
  \Sigma^{\rm LC}_{\rm soft}(\epsilon, \eta)
=
  i g^2 \mu^{2\epsilon} C_{\rm F}2 p^+ \ \int\! \frac{d^\omega q}{(2\pi)^\omega}
  \frac{1}{q^2 (q^- \cdot p^+ - i0) [q^+]_\eta} \ ,
\label{eq:soft}
\end{equation}
which contains a new singularity, that can not be circumvented by
dimensional regularization or by the $\eta$-cutoff, leading to
\begin{equation}
  \Sigma^{\rm LC}_{\rm soft}(\epsilon, \eta)
=
  - \frac{\alpha_s}{\pi} C_{\rm F}  \left[\frac{4\pi \mu^2}{\lambda^2}\right]^\epsilon
  \Gamma(\epsilon) \
  \int_0^1 dx \frac{x}{x^2 [ x-1 ]_\eta} \ ,
\label{eq:soft_int}
\end{equation}
where $\lambda$ is the IR regulator.
In our previous paper \cite{CS08}, we have argued that this divergence
is irrelevant, since it doesn't affect the rapidity evolution.
However, for the sake of completeness, we propose here a procedure,
which allows one to remove this divergence in a proper way.
Taking into account that the extra singularity is cusp-independent,
we conclude that it represents the self-energy of the Wilson line,
evaluated along a ``straightened'' path, i.e., assuming that the cusp
angle becomes very small: $p^+ \to \eta$.
Subtraction of this self-energy part is presented graphically in
Fig.\ 1.
Note that there is no need to introduce additional parameters in this
subtraction.
Moreover, it has a clear physical interpretation: only an irrelevant
contribution due to the self-energy of the light-like gauge links is
removed, which is merely part of the unobservable background.

\begin{figure}[h]
  \includegraphics[scale=0.7,angle=90]{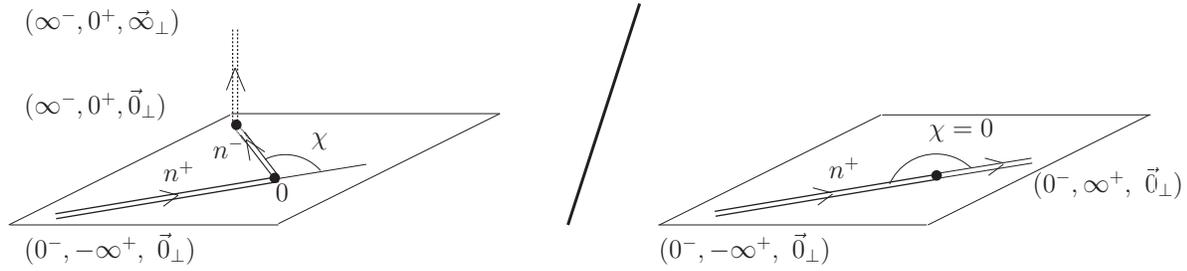}
  \caption{Subtraction of the Wilson-line self-energy contribution in
  the soft factor.}
\end{figure}

Therefore, the ``completely subtracted'' generalized definition of the
TMD reads
\begin{eqnarray}
  && {\cal F} \left(x, {\bm k}_\perp; \mu, \eta\right)
=
  \frac{1}{2}
  \int \frac{d\xi^- d^2{\bm \xi}_\perp}{2\pi (2\pi)^2}
  {\rm e}^{-ik^{+}\xi^{-} +i {\bm k}_\perp
\cdot {\bm \xi}_\perp}
  \left\langle
              h |\bar \psi_i (\xi^-, {\bm \xi}_\perp)
              [\xi^-, {\bm \xi}_\perp;
   \infty^-, {\bm \xi}_\perp]^\dagger  \right.  \nonumber \\
   && \left.
\times
   [\infty^-, {\bm \xi}_\perp;
   \infty^-, {\bm \infty}_\perp]^\dagger
   \gamma^+[\infty^-, {\bm \infty}_\perp;
   \infty^-, \mathbf{0}_\perp]
   [\infty^-, \mathbf{0}_\perp; 0^-,\mathbf{0}_\perp]
   \psi_i (0^-,\mathbf{0}_\perp) | h
   \right\rangle
    R^{-1} \ ,
   \label{eq:general}
\end{eqnarray}
\begin{eqnarray}
&& R^{-1}(\mu, \eta)
= \nonumber \\
&& \frac{\langle 0
  |   \ {\cal P}
  \exp\Big[ig \int_{\mathcal{C}_{\rm cusp}}\! d\zeta^\mu
           \ {\cal A}^\mu (\zeta)
      \Big] \cdot
  {\cal P}^{-1}
  \exp\Big[- ig \int_{\mathcal{C'}_{\rm cusp}}\! d\zeta^\mu
           \ {\cal A}^\mu (\xi + \zeta)
      \Big]
  {| 0
  \rangle } }
{ \langle 0
  |   \ {\cal P}
  \exp\Big[ig \int_{\mathcal{C}_{\rm smooth}}\! d\zeta^\mu
           \ {\cal A}^\mu (\zeta)
      \Big] \cdot
  {\cal P}^{-1}
  \exp\Big[- ig \int_{\mathcal{C'}_{\rm smooth}}\! d\zeta^\mu
           \ {\cal A}^\mu (\xi + \zeta)
      \Big]
  | 0
  \rangle  } \ ,
\end{eqnarray}
where the cusped and smooth contours are presented in Fig.\ 1.


To conclude, we have demonstrated that the generalized definition of
the TMD (\ref{eq:general}) is completely gauge- and
regularization-invariant, renormalizable and free of any kind of
emergent overlapping divergences, including those produced by the
artifacts of the soft factor --- at least in leading $\alpha_s$-order.
For completeness, one has yet to prove that this definition is part of
a TMD factorization theorem (see for an example of such an explicit
proof in covariant gauges with gauge links shifted from the
light-cone in \cite{JMY04} and the discussion in Ref. \cite{CRS07}),
and clarify the relationship of our approach (in particular, the
precise form of the soft factors, which might vary within different
schemes) to other approaches for the operator definitions of
TMDs (e.g., Refs. \cite{CM04, CH00}).
This issue is left for future work.


\paragraph{Acknowledgments}
I.O.Ch. is grateful to the Organizers of the Workshop  ``Diffraction 2010''
for the invitation and to the INFN for financial support.



\end{document}